\documentstyle[12pt,bezier]{article}
\newcommand{\OV}{\overline}
\newcommand{\MC}{\multicolumn}

\begin{document}

\title{A new family of non--linear filters for background subtraction of
       wide--field surveys}
\author{
V.S.Shergin, A.Yu.Kniazev, V.A.Lipovetsky\\
\\
Special Astrophysical Observatory of Russian Academy of Sciences\\
Nizhnij Arkhyz, Karachai-Circessia, Russia, 357147}

\maketitle
\begin{abstract}
In this paper the definitions and the properties of a newle dedicated
set of high-frequency filters based on smoothing-and-clipping are briefly
described. New applications for reduction of wide--field
2048$\times$2048 CCD spectral and direct images of a new deep survey
KISS are also presented.

\end{abstract}
\section{Introduction}

The reduction of 2D astronomical images requires a proper subtraction of
the night sky.
Even a perfect flat-field may correct only chip inhomogeneities
or optical vignetting on the CCD frames but there are many other
sources of additive noise that can not be removed with flat-fielding;
thus, there is a general problem of building a correct background which
includes such additive noise. Sometimes it is possible to use a 2D polynomial
approximation or other fittings to take into account a small gradient across
the field, but, in general, such methods do not help much, and this task can
not be solved by the usual methods incorpoprated within standard reduction
systems such as IRAF or MIDAS.

The possible sources of the additive noise include a gradient throughout
a large field ($\geq 1^\circ$) due to vicinity of the Moon, traces of light
clouds when the weather is not perfect, faults of the electronics, optical
ghosts
inside telescope optics, etc.  All these effects create large-scale
features up to the size of whole frame.
Some known extra problems rise when a Schmidt telescope and an
objective prism are used; one can see many reflections from bright stars out
of a studied field, scattered and reflected by many optical surfaces light.
Of course, they can not be removed with flat-fielding because of their additive
nature. On the other hand, an application of an adaptive filter requires
good flat frames without any gradients (Lorenz et al. 1993).
Usually a threshold is applied in order to create a correct mask
for a reasonable estimate of the noise.

In 1994, two of the authors (A.K. and V.L.) ran into these severe problems
processing data from a new survey KISS (KPNO International
Spectral Survey (Salzer et al 1994).
KISS is a newly dedicated spectral survey. It utilizes the Burrel
Schmidt telescope with a 2$^\circ$ objective prism, 2048$\times$2048 CCD,
and a special 4800$\div$5500~\AA\ filter blocking strongest lines of
the night sky.
KISS aims to search for emission-line galaxies in the region of
the Century Survey in a 1$^\circ$ slice going through the Northern Galactic
Pole.

There are several known ways how to build a background or to avoid it during
processing. They may use wavelet transformation, adaptive filtering,
median filtering, etc. (cf. Bijaoui, 1980; Slezak et al., 1990;
Coupinot et al., 1992; Lorenz et al., 1993).
Good results for removing some large-scale features had been achieved with
fast 2D median filter (and filters based on any other order statistics)
(Pasian, 1991). Nevertheless, these algorithms are very time consuming due
to 2D nature when large features should be removed.
We consider the problem of background fitting as a problem of robust
estimation of an average value in the window where objects of
interest are mixed up with all other sources from the background noise.

The authors use another approach called the smoothing-and-clipping algorithm
(SAC), considering any image as a set of independent rows or columns, 
filtering them as 1D structures, and combining them after the filtering.
Initially this algorithm had been used for reduction of 1D
radio data for the detection of weak sources during the deep survey
conducted with the RATAN--600 radiotelescope (Erukhimov, Vitkovsky and
Shergin, 1990; Verkhodanov et al. 1992, 1993) in the Special Astrophysical
Observatory. The algorithm has been modified by the authors for the application
to 2D optical images and 1D optical spectra.

For radio data a signal consists of the positive sources detected above a
certain background level plus noise. Data processing
is then a removal of such a background and the robust estimation of
the remaining sources.
There is a certain similarity between a single row (column) of CCD (with
superimposed stars and galaxies) and a radio record.
The reflections of a bright star inside the telescope optics form
a background, which is an additive signal, that should be removed when
processed.
There are some differences between optical and radio data. The latter are
always supposed to have a constant noise level.
CCD data, on the other hand,  may have a variable noise. For example,
observations with noticeable vignetting after flat-fielding
have a quite different noise over the field, and this fact has to be
taken into account to get the correct S/N value.

This approach might be applied to optical spectra to fit the optical
continuum as some sort of background. But such spectra are more complicated
due to possible emission and/or absorption lines and to the more
sophisticated shape of the real spectral continuum.
So, for optical range there is always a variable noise along
the spectrum because of variable spectral sensitivity and may be also be
other effects. They required us to modify our algorithms to include
a variable noise, their description will be publushed elsewhere,
this paper deals with the processing of 2D images.

\section{Description of the algorithm}

In general we mean not single but a family of such SAC algorithms and
all of them specified by two main steps:
smoothing and clipping, repeated iteratively.

Consider some definitions before the algorithm description:
\begin{itemize}
   \item Array S(x) is an inputed 1D spectrum or initial data (row or column);
    define it as $\OV{S}$.

   \item Let there be $\lambda$ a window width of the smoothing algorithm.
     We suppose $\lambda$ is larger than any source width (emission or
     absorption line width for a spectrum).

   \item Clipping procedure of the algorithm deals with the noise level
      $\sigma(x)$. In the general case it changes from point to point for
      input data and we define it as a vector $\overline{\sigma}$.

   \item $\overline{W}_{n}$ is a vector of weights for the n-th iteration
     smoothing. The initial value $\OV{W}_0=1$, and $\OV{W}_0$ runs in
     the range from 1 to 0.
    The weight vector shows a relative contribution of each point during
    smoothing\footnote{
    As a by-product the weight vector provides the information concerning
    the position of detected sources at the end on the SAC procedure}.
\end{itemize}

Then SAC algorithm performs following steps for every iteration:
\begin{enumerate}

\item Calculates an input vector $\OV{C}_i$ for the $\bf i-th$ iteration
   of smoothing as:
    \begin{equation}
	\overline{C}_i = {\cal F}(\overline{S},\overline{W}_{i-1})
    \end{equation}
  The index $\bf i$ is the number of the iteration.
  $\overline{W}_{i-1}$ is a vector of weights calculated in the
  previous iteration.
  In the simplest case the function $\cal F$ is calculated as a product of
  each vector component
   -- ${\cal F}_k = S_k*W_k$, where $k$ is the $k$-component.

\item Smoothes $\OV{C}_i$. Define $\Lambda_i(\lambda)$ as the smoothing
   operator.
  Index $\bf i$ shows that generally $\Lambda_i(\lambda)$ may depend on
  iteration number.
  Obviously the operator $\Lambda_i(\lambda)$ depends on the input parameter 
$\lambda$.
  The result of such an operation upon input vector $\overline{C}_i$ is:
  \begin{equation}
     \overline{B}_i = \Lambda_i(\lambda)\:\overline{C}_i
  \end{equation}
  For the first iteration, $\overline{C}_1 = \overline{S}$, according to
  definition.

\item Subtracts $\overline{B}_i$ from the input data vector $\overline{S}$:
\begin{equation}
   \overline{D}_i = \overline{S} - \overline{B}_i
\end{equation}

\item Calculates a new function of weights as a non-linear transformation
   $\Phi_i(\OV{\sigma})$, which is a function of input noise $\OV{\sigma}$
   and the number of the iteration in a general case:
   \begin{equation}
      \OV{W}_i = \Phi_i(\OV{\sigma})\:\OV{D}_i
   \end{equation}

\item Goes to the next iteration. SAC algorithm does not converge as a rule;
  thus, one should choose when to stop the procedure\footnote{
  Usually, the authors exploit between 4 and 6 iterations.}.
\end{enumerate}

As was aforementioned, SAC presents a family of filters. There are several
possible smoothing and weighting methods. In practice we explored
the following methods for smoothing:
\begin{itemize}
  \item simple boxcar average;
  \item median average (Erukhimov, Vitkovski and Shergin 1990);
  \item convolution with Gaussian--like expanding profile\footnote{
Algorithm feels width features and expand smooth--window for them};
  \item weighted least squares polynomial approximation of 1$\div$15 degrees.
\end{itemize}

To calculate the weights some empirical transformations are used. They may be
specialized to cut emission, absorption or both.
The authors employ their own empiric function plotted in Figure~1 for all
smoothing methods.
We should make two notes: a) there is a known problem of smoothing on edges;
this problem should be solved in any particular case, b) for optical data
the noise may be variable throughout the image (spectrum); thus,
our program has been modified for such a case.

Figure~2 illustrates some iterations for 1D data (a single row from KISS
CCD data).
It is clear that the result the after 5-th iteration is quite satisfied.

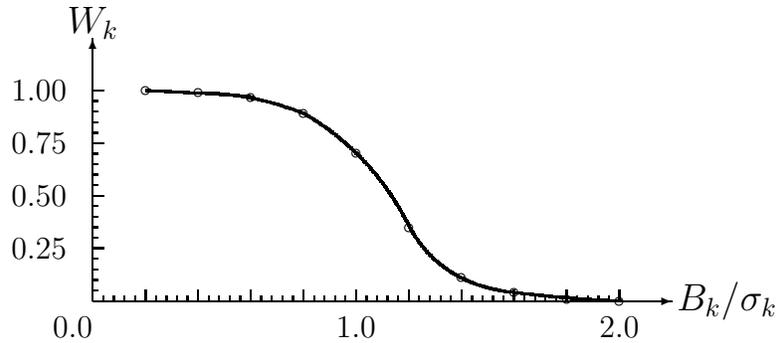
\begin{figure}
     \begin{center}
\begin{picture}(12,6.5)

\put(1,1){\vector(1,0){11}}
\multiput(1,1)(1,0){11}{\line(0,1){.2}}
\multiput(1,1)(0.2,0){50}{\line(0,1){.1}}
\put(1,0.5){\makebox(0,0)[r]{0.0}}
\put(6,0.5){\makebox(0,0){1.0}}
\put(11,0.5){\makebox(0,0){2.0}}
\put(12.1,1){\makebox(0,0)[l]{\large $B_k/\sigma_k$}}

\put(1,1){\vector(0,1){5}}
\multiput(1,2)(0,1){4}{\line(1,0){.2}}
\multiput(1,1)(0,0.2){20}{\line(1,0){.1}}
\put(0.5,2){\makebox(0,0)[r]{0.25}}
\put(0.5,3){\makebox(0,0)[r]{0.50}}
\put(0.5,4){\makebox(0,0)[r]{0.75}}
\put(0.5,5){\makebox(0,0)[r]{1.00}}
\put(1,6.1){\makebox(0,0)[b]{\large $W_k$}}

\put(2,5){\circle{0.15}}
\put(3,4.96){\circle{0.15}}
\put(4,4.88){\circle{0.15}}
\put(5,4.56){\circle{0.15}}
\put(6,3.8){\circle{0.15}}
\put(7,2.4){\circle{0.15}}
\put(8,1.44){\circle{0.15}}
\put(9,1.16){\circle{0.15}}
\put(10,1.04){\circle{0.15}}
\put(11,1.00){\circle{0.15}}

\thicklines
\bezier{100}(2,5)(3,4.96)(4,4.88)
\bezier{100}(3,4.96)(4,4.95)(5,4.56)
\bezier{150}(5,4.56)(6.3,3.8)(7,2.4)
\bezier{150}(7,2.4)(7.6,1.30)(9,1.16)
\bezier{100}(9,1.16)(10,1.04)(11,1)
\end{picture}

     \caption{Empirical transformation (weighting) used by authors. Abscissa
	    value is a difference between initial data and background
	    estimation for the {\bf i-th} iteration in terms of input noise
	    for the {\bf k-th} element of the vector.}
     \end{center}
\end{figure}

\begin{figure}
 \setlength{\unitlength}{1.0cm}
 \centering{
 \hspace*{-1.0cm}
 \vbox{\includegraphics{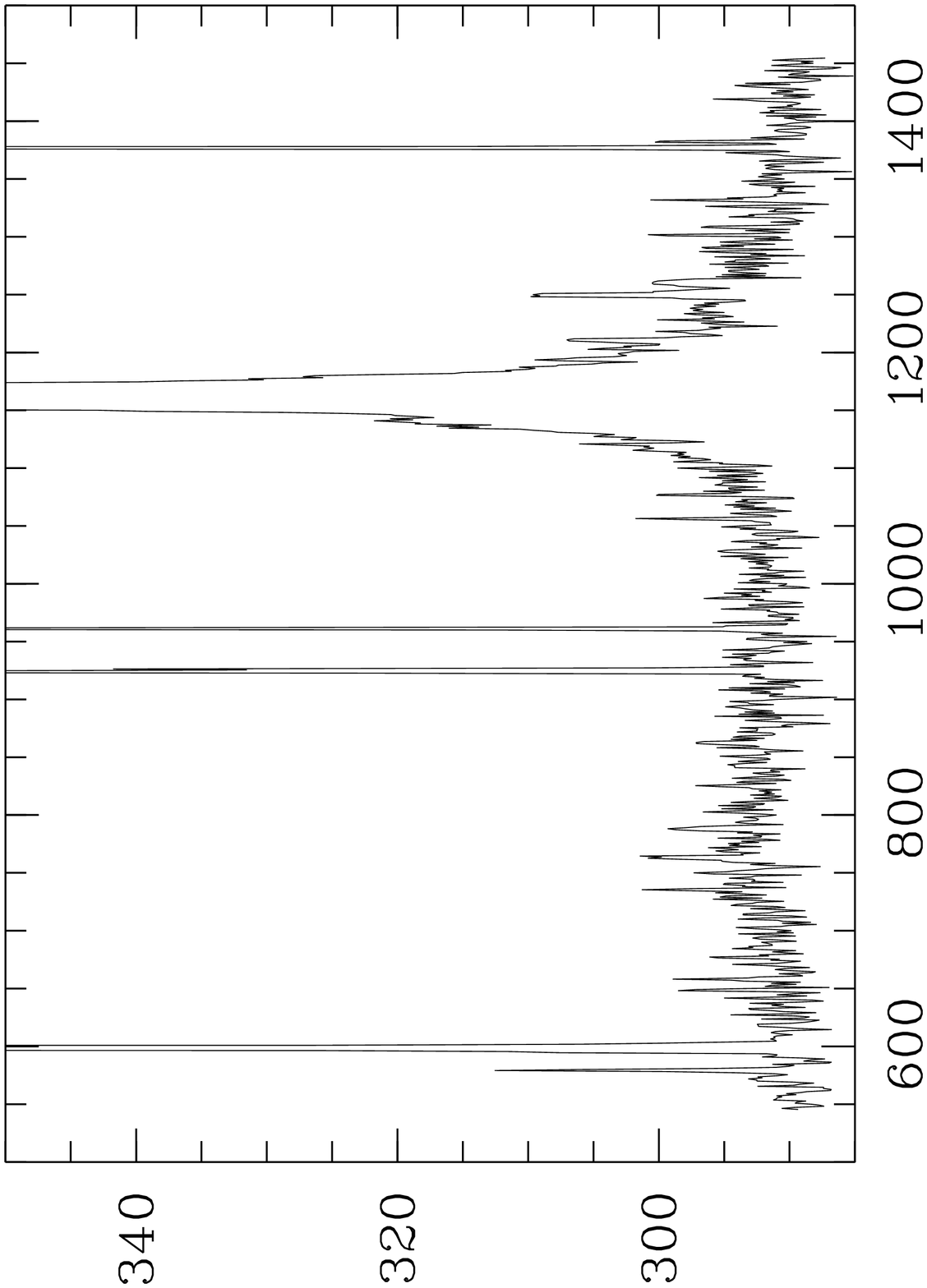}
       \includegraphics{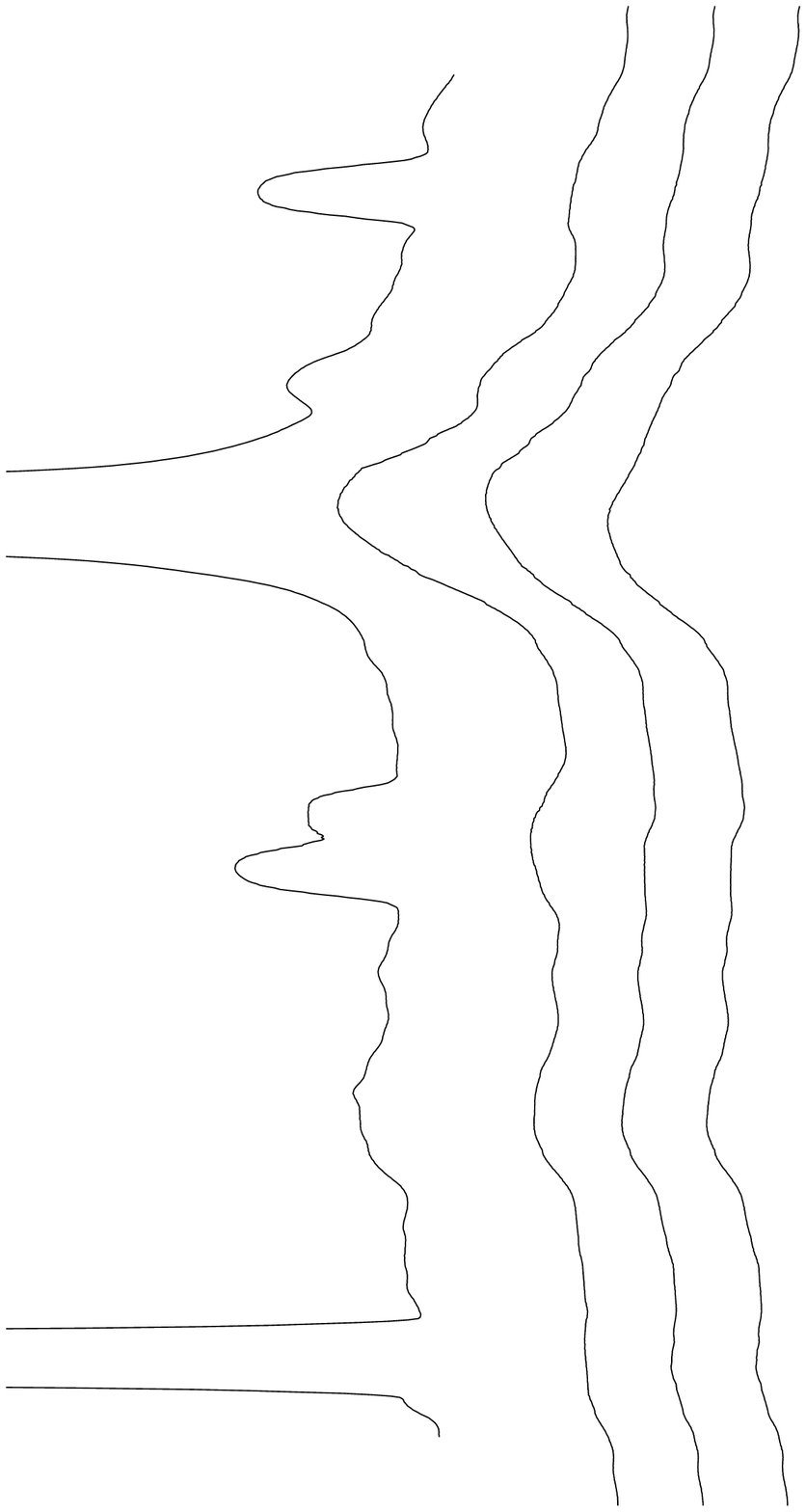}}\par
 \begin{picture}(0,0)
    \put(-6.4,-6.5){\line(1,0){12.2}}
    \put(-6.4,-6.5){\line(0,1){4.3}}
    \put(-6.4,-2.2){\line(1,0){12.2}}
    \put(5.8,-6.5){\line(0,1){4.3}}
    \put(-4.65,2.7){\line(1,0){0.65}}
    \put(-4.65,2.6){\line(0,1){0.2}}
    \put(-4.0,2.6){\line(0,1){0.2}}
    \put(-4.3,2.9){\makebox(0,0){$\lambda$}}
 \end{picture}
 \vspace*{7.0cm}
 }
     \caption{An example of SAC filter with a Gaussian smoothing window,
	      $\lambda=60$ pixels (the bar in the up--left corner shows the
	      smoothing window size). The upper figure is a portion of an
	      extracted CCD row in pixels, the bottom figure shows the 1-st
	      (simple convolution of the data with Gaussian profile), 2-nd,
	      5-th and 10-th iterations, all plots shifted arbitruary
	      along the ordinate axis. It is evident that after the 5-th
	      iteration the results practically do not change.
	      All features smaller than a window size are smoothed very
	      quickly.
	      The bright saturated star, chosen on the upper right part of
	      Figure 3, has a profile width somewhat larger
	      than $\lambda$ and te algorithm clips only the narrow part of
	      this profile}
\end{figure}
\section{Application and tricks}

The developed software is available both as a C subroutine and as an installed
MIDAS environment command (see Appendix).

This software is used to process KISS CCD direct images and taken with an
objective prism.
Figure~3 presents a quite successful result of SAC cleaning for a direct image.
It is evident that all imperfections --- such as noticeable jump of the background
due electronic "gliters" during CCD read-out, the halos of bright stars and
a gradient in the low left corner pof the frame were corrected to an accuracy
better of greater than 0.1\%.

The authors tested the program with different platforms. Their experience
showed that, thanks to the 1D approach, SAC may be used even with small
PC computers installed with MIDAS. The total time does not depend on
the window size for simple boxcar average, and the typical time is only
about a few minutes for a 2K$\times$2K frame\footnote{
In their tests the authors assumed both input and output images could be
located in computer RAM; so for a PC computer, this time depends crucially
on RAM to prevent swapping.}.
For Gaussian smoothing, the time depends strongly on window size $\lambda$.
Table~1 sums up typical times for two sizes, 2K$\times$2K images, and 3
different computers. For polynomial smoothing, the procedure time is about
4 times quicker than for Gaussian.

\begin{table}[btp]
  \caption{Typical time of SAC performance on different platforms}
  \begin{center}
   \begin{tabular}{llll}\hline\hline
    \MC{1}{c}{Platform}   &\MC{1}{c}{RAM}      & \MC{1}{c}{Time for} & 
\MC{1}{c}{Time for} \\
	& \MC{1}{c}{size} &\MC{1}{c}{$\lambda=60$ pixels}&\MC{1}{c}{$\lambda=90$ 
pixels}\\
    \MC{1}{c}{(1)}        & \MC{1}{c}{(2)}     & \MC{1}{c}{(3)} & 
\MC{1}{c}{(4)}   \\ \hline
     SUN--server NOAO     & 128 Mb &   55 min  &     2.5 hours      \\
     Sparc--20            & 64 Mb  &   45 min  &     2 hours        \\
     Convex               & 900 Mb &   2 hours &     5 hours        \\  \hline
   \end{tabular}
  \end{center}
  \label{}
\end{table}

Some tricks of SAC perfomance include:

1. The fastest algorithm is boxcar averaging but this method is
  the roughest. The compromise may be reached using boxcar averaging
  as a start and Gaussian smoothing in the end. The polinomial smoothing
  has a intermediate quality comparing with the others.

2. For a 2D frame there are two sizes for a source; thus, one may chose
  the direction, X or Y, to smooth the image. For instance, objective prism
  spectral images of stellar objects with have typical sizes 5x45 pixels;
  thus, it may be recommended to smooth across the dispersion.

3. When there are some artifacts like spikes of
  bright stars, it is better to select the dominant direction to remove
  them. In a unique case, such as tracks of a satellite or a plane,
  we recommend to rotate a frame along such tracks and then apply SAC.

4. The program needs only an innitial estimate of the noise, not a precise
   value. The usefull of this feature was shown in Figure~3 where different
   image parts has different amount of noise (difference was about 20\%).

The authors (A.K. and V.L.) are grateful to the staff of the Astrophysical
Institute Potsdam for their hospitality and the possibility to use their
Convex computer to process KISS data.
\begin{figure}[hbtp]
 \setlength{\unitlength}{1.0cm}
 \centering{
 \hspace*{-1.0cm}
 \vspace*{22.0cm}
 }
    \caption{KISS direct image for field $\alpha = 14^h45^m$ before (upper)
    and after (lower) SAC cleaning.
    The small jump on the upper image of about 1.3\% was caused by
    an elecronics failure during read--out.
    The noise value is a little different for two parts of initial image
    ($\sigma_{up}$=6.49, $\sigma_{down}$=7.78), but the SAC does not need
    it to be precise. For an initial value $\sigma_{aver}$ was taken for
    the SAC program.}

\end{figure}
%

\appendix
\section{MIDAS procedure of SAC}
The command works both for versions 2D image and for real 1D spectra\footnote{
The C code and MIDAS version may be requested from the authors through e-mail:
akn@sao.stavropol.su}.
For 2D images the parameter for spectrum type should be "both".
The working version of this command had been installed in SAO on PC computers
and Sparc 20 and in AIP on the Convex machine.

\begin{verbatim}
@s contin  inp_image  out_image  smooth_wnd  noise_level
	   iter_numb  bgd_and_out_type  smooth_mode  smooth_step

where:     inp_image   - input frame
	   out_image   - output frame
	   smooth_wnd  - width of smoothing window
	   noise_level - noise estimation
	   iter_numb   - number of iterations
	   bgd_and_out_type - two characters (default - "bb") :
	      bgd_type   -   a/e/b (for "absorbtion" or "emission" or "both")
	      out_type   -   r - residuals    b - background itself
	   smooth_mode   -  <0 - quick-rough mode, boxcar averaging
			    =0 - slow mode, convolution with Gaussian window
			    >0 - fitting polynom power (< 15)
	   fitting_step-  step of fitting polynom locations (< smooth_wnd/2)

	   Type "help" as the first parameter to get short help

\end{verbatim}

\end{document}